\documentclass[prx, aps,twocolumn,superscriptaddress,longbibliography]{revtex4-1}
\usepackage{latexsym}
\usepackage{graphicx}
\usepackage{multirow}

\newcommand{\be}{\begin{equation}}
\newcommand{\ee}{\end{equation}}
\newcommand{\ba}{\begin{eqnarray}}
\newcommand{\ea}{\end{eqnarray}}
\newcommand{\baa}{\begin{eqnarray*}}
\newcommand{\eaa}{\end{eqnarray*}}

\def\be{\begin{equation}}
\def\ee{\end{equation}}
\def\bea{\begin{eqnarray}}
\def\eea{\end{eqnarray}}

\def\C60{A$_x$C$_{60}$}

\def\HgCu3{HgCa$_2$Cu$_3$O$_{8+y}$}
\def\HgCu4{HgBa$_2$Ca$_3$Cu$_4$O$_{10+y}$}
\def\TlCu{Tl$_2$Ba$_2$CuO$_{6+\delta}$}
\def\TlCu3{Tl$_2$Ba$_2$Ca$_2$Cu$_3$O$_{10+y}$}
\def\TlCu4{Tl$_2$Ba$_2$Ca$_3$Cu$_4$O$_{12+y}$}

\def\BiCu3{Bi$_2$Sr$_2$Ca$_{2}$Cu$_3$O$_y$}

\def\8LSCO{La$_{1.88}$Sr$_{.12}$CuO$_4$}
\def\110LNSCO{La$_{1.5}$Nd$_{0.4}$Sr$_{0.1}$CuO$_{4}$}
\def\stage4LCO{La$_{2}$CuO$_{4+\delta}$}
\def\Y248{YBa$_2$Cu$_4$O$_8$}

\def\NbSe2{NbSe$_2$}
\def\TaSe2{TaSe$_2$}
\def\TiSe2{TiSe$_2$}

\begin{document}

\title{$S_4$ Symmetric Microscopic Model  for Iron-Based Superconductors}

\author{Jiangping Hu}
\affiliation{Beijing National
Laboratory for Condensed Matter Physics, Institute of Physics,
Chinese Academy of Sciences, Beijing 100080,
China}
\affiliation{Department of Physics, Purdue University, West
Lafayette, Indiana 47907, USA} 
\author{Ningning Hao}
\affiliation{Beijing National
Laboratory for Condensed Matter Physics, Institute of Physics,
Chinese Academy of Sciences, Beijing 100080,
China}
\affiliation{Department of Physics, Purdue University, West
Lafayette, Indiana 47907, USA}

\begin{abstract}
Although iron-based superconductors are multi-orbital  systems with complicated band structures, we demonstrate that the low energy physics which is responsible for high-$T_c$ superconductivity is essentially governed by an effective two-orbital Hamiltonian  near half filling. This underlining electronic structure is protected by the $S_4$  symmetry. With repulsive  or strong next nearest neighbor antiferromagnetic exchange interactions, the model results in a robust  $A_{1g}$ $s$-wave pairing which can be exactly mapped to the $d$-wave pairing observed in cuprates. The classification of the superconducting(SC) states according to the $S_4$ symmetry  leads to a natural prediction of the existence of two different phases named A and B phases. In the B phase,    the superconducting order has an overall sign change along c-axis between the top  and bottom As(Se) planes  in a single Fe-(As)Se trilayer structure, which is an analogy of the sign change under the $90^\circ$ degree rotation  in the $d$-wave SC state of cuprates.    Our derivation provides a unified understanding of iron-pnictides and iron-chalcogenides, and suggests that cuprates and iron-based superconductors share identical high-$T_c$ superconducting mechanism.  
\end{abstract}

\maketitle

\section{Introduction}
  Since the discovery of iron-based superconductors\cite{Hosono,ChenXH,nlwang,ChenXL}, there has been considerable  controversy over the choice of the appropriate microscopic Hamiltonian\cite{john,hirschfeld}.  The major  reason  behind such a controversy is  the complicated multi $d$-orbital electronic structure of the materials.   Although the electronic structure has been modeled  by using different numbers of orbitals, ranging from  minimum two orbitals\cite{raghu}, three orbitals\cite{patrick},  to all five d orbitals\cite{mazin,kuroki},  a general perception has been that any microscopic model composed of less than all five  $d$-orbitals and  ten bands is insufficient\cite{hirschfeld}. Such a perception has blocked the path to understand  the superconducting mechanism because of  the difficulty in  identifying  the key physics   responsible for high $T_c$. 
Realistically, in a model with five orbitals, it is very difficult for any theoretical calculation to make meaningful predictions in a controllable manner. 

Iron-based superconductors include two families, iron-pnictides\cite{Hosono,ChenXH,nlwang} and iron-chalcogenides\cite{ChenXL}.  They share many intriguing common properties. They both have the highest $T_c$s   around 50K\cite{john,ChenXH,liud,qwang,lsun}. The superconducting gaps  are close to isotropic   around Fermi surfaces\cite{hding, zhouxj, zhangy2,Wang_122Se, Zhang_122Se,Mou_122Se}  and the ratio between the gap  and $T_c$, $2\Delta/T_c$,   are much larger than the BCS ratio, 3.52, in both families.  However, the electronic structures in the two  families, in particular, the Fermi surface topologies, are quite different in the materials  reaching  high $T_c$.   The hole pockets are absent in iron-chalcogenides but present in iron-pnictides\cite{hding,Wang_122Se, Zhang_122Se,Mou_122Se}. The presence of the hole pockets has been a necessity for superconductivity in  the majority of studies and models which deeply depend on the properties of  Fermi surfaces. Therefore, the absence of the hole pockets in iron-chalcogenides causes a strong debate over whether both families belong to the same category that shares a common superconducting mechanism.  Without  a clear microscopic picture of the underlining electronic structure, such  a debate can not be  settled.

Observed by angle-resolved photoemission microscopy (ARPES),  a very intriguing property in the SC states of iron-pnictides is that the SC gaps on different Fermi surfaces are nearly proportional to a simple form factor $cos k_x cosk_y$ in reciprocal space. This form factor has been observed in both  $122$\cite{hding, hding2, zhouxj,Nakayama}  and $111$\cite{umezawa,liuzh} families of iron-pnictides.  Just like the $d$-wave form factor $cosk_x-cosk_y$ in cuprates, such a form factor indicates that  the pairing between two  next nearest neighbour iron sites in  real space dominates.  In a multi orbital model, many theoretical calculations based on weak coupling approaches have shown that the gap functions are very sensitive to detailed band structures and  vary significantly when  the doping changes\cite{hirschfeld, WangF, thomale1, thomale2, chubukov,zlako}. The robustness of the form factor has been  argued to favor strong coupling approaches which emphasize  electron-electron correlation or the effective  next nearest neighbour (NNN) antiferromagnetic (AF) exchange coupling $J_2$\cite{seo2008,Fang2011,local1, hu1,hu4,luxl,berg} as a primary source of  the pairing force.   
However, realistically, it is very difficult to imagine such a local exchange interaction remains  identical between all $d$-orbital electrons if  a multi $d$-orbital model is considered.

In this paper, we demonstrate that the underlining electronic structure in iron-based superconductors, which is responsible for superconductivity at low energy,  is essentially governed by a two orbital model obeying the $S_4$ symmetry. The two orbital model includes two nearly degenerated single-orbital parts that can be mapped to each other under the $S_4$ transformation.    This electronic structure  stems from  the fact that the dynamics of  $d_{xz}$ and $d_{yz}$ orbitals  are divided into two groups that are separately coupled to the top and bottom As(Se) planes in a in a single Fe-(As)Se trilayer structure. The two groups can thus be treated as a $S_4$ iso-spin. The dressing of other orbitals in the $d_{xz}$ and $d_{yz}$ orbitals can not alter the symmetry characters.

\begin{figure}[tbp]
\begin{center}
\includegraphics[width=1.0\linewidth]{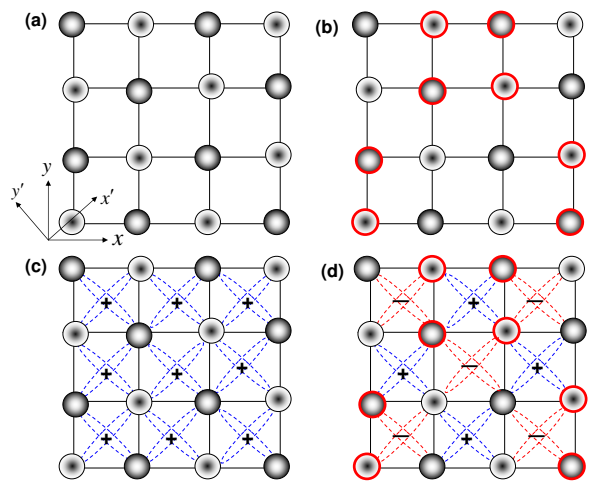}
\end{center}
\caption{ (a) the square lattice structure of a single iron layer: one cell includes two Fe ions identified with different filled black balls which form two sublattices. We use $x-y$ coordinate to mark the original tetragonal lattices and $x'-y'$ to mark the sublattice direction. (b) the gauge transformation is illustrated. The balls with red circles   are affected by the gauge transformation.  (c) and (d) the mapping from the $s$-wave to the $d$-wave pairing symmetry by the gauge transformation.}
\label{fig1}
\end{figure}

The underlining electronic structure becomes transparent after performing a gauge mapping in the five orbital model\cite{kuroki}.  The gauge mapping also reveals the equivalence between the $A_{1g}$ $s$-wave pairing  and the $d$-wave pairing. After the gauge mapping, the band structure for each $S_4$ iso-spin component   is characterized by   Fermi surfaces   located around the anti $d$-wave nodal points in Brillouin zone, corresponding to the sublattice periodicity of the bipartite iron square lattice as shown in Fig.\ref{fig1}(a).  In the presence of an AF exchange coupling $J_2$ or an effective on-site Hubbard  interaction,   the $d$-wave pairing defined in the sublattices   can be argued to be favored,  just like  the case in cuprates.   The $d$-wave pairing symmetry   maps reversely  to a $A_{1g}$ $s$-wave pairing  in the original gauge setting.  These results provide a unified microscopic understanding of iron-pnictides and iron-chalcogenides and  explain why an $s$-wave  SC state without the  sign change on Fermi surfaces  in iron chalcogenides driven by repulsive interaction  can be so robust.   More intriguingly,  since the different gauge settings do not alter any physical measurements,  the results suggest  that in  the $A_{1g}$ $s$-wave  state, for each $S_4$ iso-spin component, there is a hidden sign change between the top As(Se)  and the bottom As(Se) planes along c-axis. 

The $S_4$ symmetry adds a new symmetry classification to the SC states. For example, even in the  $A_{1g}$ $s$-wave pairing state, there are two different phases called $A$ and $B$ phases, with respect to the $S_4$ symmetry. In the $A$ phase, the relative SC phase between the two $S_4$ iso-spin components is zero while in the $B$ phase, it is $\pi$. Therefore,  there is an overall $\pi$ phase shift between the top As(Se)  and the bottom As(Se) planes in the B phase along the c-axis. Such a sign change should be detectable experimentally.  This property   makes iron-based superconductors  useful in many SC  device applications. An experimental setup, similar to those for determining the $d$-wave pairing in cuprates\cite{woll,van,tsuei},  is proposed to detect the $\pi$ phase shift.  The detection of the sign change  will strongly support that cuprates and iron-based superconductors share identical microscopic superconducting mechanism and will establish that  repulsive interactions are  responsible for superconductivity.
 
The paper is organized in the following way.  In Section II, we perform a gauge mapping  and discuss the emergence of the  underlining electronic structure. In Section III,
we show that the underlining electronic structure can be constructed by a  two orbital model obeying the $S_4$ symmetry and discuss many general properties in the model. In Section IV, we discuss the classification of the SC states under the $S_4$ symmetry and propose a measurement to detect the $\pi$ phase shift along c-axis between the top and bottom As(se) planes.  In Section V,   we discuss the analogy between iron-based superconductors and cuprates.

\section{Gauge mapping and the equivalence of $s$-wave and $d$-wave pairing}
{\it Gauge Mapping:}   We start to ask whether there is an unidentified  important electronic structure in iron-based superconductors in a different gauge setting. Giving a translational invariant Hamiltonian  that describes the electronic band structure of  a Fe square lattice,
\be
\hat H_0=\sum_{ij,\alpha\beta,\sigma} t_{ij,\alpha\beta}\hat f_{i\alpha,\sigma}^+\hat f_{j\beta,\sigma} ,
\ee
where $i,j$ label Fe sites,  $\alpha,\beta$ label orbitals and $\sigma$  labels spin.  We consider the following gauge transformation.  As shown in Fig.\ref{fig1}(a,b), we group four neighbouring iron sites to form a super site  and mark   half super sites by red color.  The  gauge transformation,  $\hat U$,   adds a minus sign to all Fermionic operators $\hat f_{i\alpha,\sigma} $ at every site $i$  marked by  red color.   After the transformation, the Hamiltonian becomes
\be 
 \hat H'_0=\hat U^+\hat H_0\hat U.  
\ee
The gauge mapping operator $\hat U$ is an unitary operator so that the eigenvalues of $\hat H_0$ are not changed after the gauge transformation.   It is  also important to notice that the mapping does not change standard  interaction terms, such as  conventional  electron-electron interactions  and  spin-spin exchange couplings. Namely, for a general Hamiltonian including  interaction terms $\hat H_I$, under  the mapping, 
\be 
 \hat H= \hat H_0+\hat H_I \rightarrow \hat H'= \hat U^+\hat H\hat U=  \hat H'_0+\hat H_I. 
\ee

It is also easy to see that every unit cell of the lattice in the new gauge setting includes four iron sites. The original translational invariance of a Fe-As(Se) layer has  two Fe sites per unit cell.  As we will show in the following section, the doubling of the unit cell matches the true hidden unit cell in the electronic structure when the orbital degree of freedom is considered.  This is the fundamental reason that the new gauge happens to reveal the underlining electronic structure.

{\it Equivalence of $s$-wave and $d$-wave pairing:} The gauge mapping has another important property.  As shown in Fig.\ref{fig1}(c,d), 
this transformation maps the $A_{1g}$ $s$-wave $cos(k_x)cos(k_y)$ pairing symmetry in the original Fe lattice to a familiar $d$-wave $cosk'_x -cosk_y'$ pairing symmetry defined in the two sublattices, where $(k_x,k_y)$ and $(k_x',k_y')$ label momentum in Brillouin zones of the origin lattice and  sublattice respectively.  A similar mapping has been discussed in the study of a two-orbital iron ladder model\cite{berg,berg2} to address the equivalence of $s$-wave and $d$-wave pairing symmetry in one dimension. 

In an earlier paper\cite{hu1},  one of us and his collaborator suggested  a phenomenological necessity for achieving  high $T_c$ and  selecting pairing symmetries:  when the pairing is driven by a local AF exchange coupling,  the pairing form factor has to match the Fermi surface topology  in reciprocal space.  If this rule is valid and the iron-based superconductors are in the $A_{1g}$ $s$-wave state, we   expect that the Fermi surfaces after the gauge mapping should be located in the   $d$-wave anti-nodal points in the sublattice Brillouin zone, which is indeed the case as we will show in the following.

\begin{figure}[tbp]
\begin{center}
\includegraphics[width=1.0\linewidth]{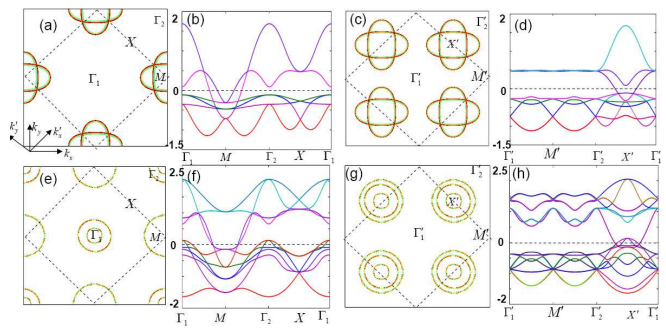}
\end{center}
\caption{   (Three\cite{Fang2011}, five\cite{kuroki})  orbital models: (a,e)  the Fermi surfaces, (b,f) the band dispersion along the high symmetry lines, (c,g) the Fermi surfaces after the gauge transformation, (d,h) the band dispersions along the high symmetry lines after the gauge transformation.  The  hopping parameters can be found in the above two references.}\label{fig2}
\end{figure}

{\it Band structures after gauge mapping:}
 There have been various tight binding models to represent the band structure of $\hat H_0$.  In Fig.\ref{fig2},  we plot the band structure of $\hat H_0$ and the corresponding  $\hat H'_0$ for  two different  models:  a maximum five-orbital model for iron-pnictides\cite{kuroki}, and a three-orbital model constructed for electron-overdoped iron-chalcogenides\cite{Fang2011}.  

As shown in Fig.\ref{fig2},  although there are subtle differences  among the band structures of $H_0'$, striking common features are revealed  for both models. First, exactly as expected,  all Fermi surfaces after the gauge mapping  are relocated around $X'$,  the  anti-nodal points in  a standard $d$-wave superconducting state  in the sublattice Brillouin zone. This is remarkable because a robust $d$-wave superconducting state can be argued to be favored  in such a Fermi surface topology in the presence of repulsive interaction or nearest neighbour (NN) AF coupling in the sublattice\cite{scalapino,hu1}.  If we reversely map to the original gauge,  the original Hamiltonian must have a robust $s$-wave pairing symmetry. Therefore, an equivalence between the $A_{1g}$ $s$-wave  and   the  $d$-wave pairing is clearly   established by the gauge mapping. 

Second,  the bands previously located at the different places on the  Fermi surface are magically linked in the new gauge setting.  In particular,    the two bands that contribute to electron pockets are nearly degenerate and in the five orbital model, the bands that contribute to hole pockets are remarkably connected to them.
Considering the fact that  the unit cell has four iron sites in the new gauge setting,   this unexpected connections lead us to believe that in the original gauge,  there should be  just
two orbitals  which form bands that make connections from  lower energy bands to higher energy ones and determine Fermi surfaces. Moreover,  the two orbitals should form two groups which provide  two nearly degenerate band structures.
 Finally,   since the mapping does not change electron density,    Fig.\ref{fig2} reveals the doping level in each  structure  should be  close to half filling.  

In summary, the gauge  mapping reveals  that the low energy physics is controlled by a two orbital model that produces two nearly degenerated bands.

\section{The construction of a two-orbital model with the $S_4$ symmetry}
With above observations,  we move to construct an effective two orbital model to capture the underlining electronic structure revealed by the gauge mapping. 

{\it Physical picture:}  Our construction is guided by the following several facts. First,  the $d$-orbitals that form the bands near the Fermi surfaces are strongly hybridized with the p-orbitals of As(Se). Since the $d_{x'z}$ and $d_{y'z}$ have the largest overlap with the $p_{x'}$ and $p_{y'}$ orbitals, it is natural for us to use $d_{x'z}$ and $d_{y'z}$ to construct the model. Second, in the previous construction of a two-orbital model, the $C_{4v}$ symmetry was used\cite{raghu}. The $C_{4v}$ symmetry is not a correct symmetry if the hopping parameters are generated through the p-orbitals of $As(Se)$. Considering the $As(Se)$ environment, a correct symmetry for the $d$-orbitals at the iron-sites is  the $S_4$ symmetry group. Third, there are two As(Se) planes which are separated in space along c-axis. Since there is little coupling between the $p$ orbitals of the two planes and the hoppings through the p-orbitals  are expected to dominate over the direct exchange hoppings between the $d$-orbitals themselves,
 the two orbital model essentially could 
be decoupled into two nearly degenerated one orbital models.  Finally, the model should have a translational invariance with respect to the As(Se) plane. 
\begin{figure}[tbp]
\begin{center}
\includegraphics[width=1.0\linewidth]{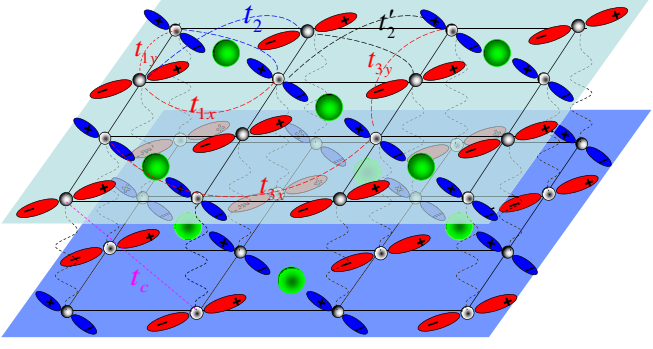}
\end{center}
\caption{  A sketch of the $ d_{x'z}$ and $ d_{y'z}$ orbitals, their  orientations and their coupling  into the two As(Se) layers.  The hopping parameters are indicated: the nearest neighbor hopping is marked by $t_{1x,1y}$, the next nearest neighbor hoppings are $t_{2}$ and $t_{2}^{\prime }$ due to the broken symmetry along two different diagonal directions, the third NN hopping is marked by $t_{3x,3y}$. The coupling between two layers is marked  by the nearest neighbor hopping $t_{c}$.}
\label{fig3}
\end{figure}

With above guidelines,  it is very natural for us to divide  the two $d$-orbitals   into two groups as shown in Fig.\ref{fig3}. One group includes the $d_{x'z}$  in the A sublattice and the $d_{y'z}$ in the B sublattice, and the other includes the $d_{x'z}$  in the B sublattice and the $d_{y'z}$ in the A sublattice,
 where A and B label the two sublattices of the iron square lattice as shown in Fig.\ref{fig1}(a).  The first group strongly couples to the p-orbitals in the up As(Se) layer and  the second group couples to those in the bottom As(Se) layer.  We denote  $\hat c_{i\sigma}$ and $\hat d_{i\sigma}$  as Fermionic operators for the two groups respectively at each iron site.

{\it $S_4$ symmetry and the two-orbital model:} Without turning on couplings between the two groups, we seek a general tight binding model to describe the band structure based on the $S_4$ symmetry. The $S_4$ transformation maps   $\hat c_{i\sigma}$ to   $\hat d_{i\sigma}$.  If we define the corresponding operators in  momentum space as $\hat c_{k\sigma}$ and $\hat d_{k\sigma}$, the $S_4$ transformation takes
\begin{eqnarray}
\left( \begin{array}{c}
\hat c_{k\sigma}\\
\hat d_{k\sigma}
\end{array}\right) \rightarrow \left( \begin{array}{c}
-\hat d_{k'+Q\sigma}\\
\hat c_{k'+Q\sigma}
\end{array}\right),
\end{eqnarray}   
where $k'=(k_y,-k_x)$ and $Q=(\pi,\pi)$ for given $k=(k_x,k_y)$.

Now, we consider a tight binding model for the first group.  Here we limit  the hopping parameters up to the third NN (TNN). As illustrated in Fig.\ref{fig3}, the tight binding model can be approximated by including NN hoppings, $t_{1x}$, $t_{1y}$,   NNN hoppings,  $t_2$, $t_2'$, and TNN hoppings, $t_{3x}$ and $t_{3y}$. The longer range hoppings can be included  if needed.  For convenience, we can define $t_{1s}=(t_{1x}+t_{1y})/2$, $t_{1d}=(t_{1x}-t_{1y})/2$, $t_{2s}=(t_2+t'_2)/2$ and $t_{2d}=(t_2-t_2')/2$, $t_{3s} = (t_{3x}+ t_{3y})/2$ and  $t_{3d} = (t_{3x}- t_{3y})/2$, where the labels, $s$ and $d$, indicate  $s$-wave    (hopping parameter  is symmetric  under  the $90^\circ$ degree rotation)  and $d$-wave (hopping parameter  changes sign  under  the $90^\circ$ degree rotation) type hoppings respectively.   
A general tight binding model can be written as
\begin{widetext}
\begin{eqnarray}
  & & \hat H_{0,one}= \sum_{k,\sigma} 2[t_{1s}(cosk_x+cosk_y)-\frac{\mu}{2} 
  + t_{1d}(cosk_x-cosk_y)]\hat c^+_{k\sigma}\hat c_{k\sigma}  
  + 4[t_{2s}cosk_xcosk_y\hat c^+_{k\sigma}\hat c_{k\sigma}+t_{2d}sink_xsink_y\hat c^+_{k\sigma}\hat c_{k+Q\sigma}] \nonumber \\
& & + 2 [t_{3s}(cos2k_x+cos2k_y) +t_{3d}(cos2k_x-cos2k_y)]\hat c^+_{k\sigma}\hat c_{k\sigma}  
  +... 
\end{eqnarray}.
\end{widetext}
We can apply the $S_4$ transformation on $\hat H_{0,one}$ to obtain the tight binding model for the second group.  The transformation invariance requires $t_{1s}, t_{2d}$ and $t_{3d}$ to change signs.  Therefore, the two-orbital model is described by
\begin{eqnarray}
  \hat H_{0,two} & =& \sum_{k\sigma} [4t_{2s} cosk_xcosk_y -\mu](\hat c^+_{k\sigma}\hat c_{k\sigma}+\hat d^+_{k\sigma}\hat d_{k\sigma})\nonumber \\
& & +2t_{1s}(cosk_x+cosk_y)(\hat c^+_{k\sigma}\hat c_{k\sigma}-\hat d^+_{k\sigma}\hat d_{k\sigma}) \nonumber \\
& & +2t_{1d}(cosk_x-cosk_y)(\hat c^+_{k\sigma}\hat c_{k\sigma}+\hat d^+_{k\sigma}\hat d_{k\sigma}) \nonumber \\
& &+4t_{2d} sink_xsink_y (\hat c^+_{k\sigma}\hat c_{k+Q\sigma}-\hat d^+_{k\sigma}\hat d_{k+Q\sigma}) \nonumber \\
& &+2t_{3s}(cos2k_x+cos2k_y)(\hat c^+_{k\sigma}\hat c_{k\sigma}+\hat d^+_{k\sigma}\hat d_{k\sigma}) \nonumber \\
& & +2t_{3d}(cos2k_x-cos2k_y)(\hat c^+_{k\sigma}\hat c_{k\sigma}-\hat d^+_{k\sigma}\hat d_{k\sigma}) \nonumber \\
& & +... 
\label{hh}
\end{eqnarray}
Now we can turn on the couplings between the two groups.   It is straightforward to show that the leading order of the couplings that satisfies the $S_4$ symmetry is given by 
\begin{eqnarray}
\hat H_{0,c}=\sum_k 2 t_c (cosk_x+cosk_y)(\hat c^+_{k\sigma}\hat d_{k\sigma}+h.c.).
\label{tc}
\end{eqnarray} 
Combining the $\hat H_{0,two}$ and $\hat H_{0,c}$, we obtain  an effective  $S_4$ symmetric two orbital model whose band structure is described by
\begin{eqnarray}
\hat H_{0,eff}= \hat H_{0,two}+\hat H_{0,c}.
\end{eqnarray}
The $\hat c$ and $\hat d$ Fermionic operators can be viewed as two  iso-spin components of the $S_4$ symmetry.


Let's assume $t_c$ to be small and check whether $H_{0,eff}$ can capture the electronic structure at low energy. Ignoring $t_c$,  $H_{0,eff} $ provides the following energy dispersions for the two orbitals, 
\begin{widetext}
 \begin{eqnarray}
& & E_{e\pm}=\epsilon_k \pm
 2 t_{3d} (cos2k_x - cos2k_y) +    4 \sqrt{t_{2d} ^2 sin^2x sin^2y  + [\frac{t_{1s} (cosk_x + cosk_y) \pm t_{1d} (cosk_x - cosk_y)}{2}]^2},\\
& & E_{h\pm}=\epsilon_k \pm
 2 t_{3d} (cos2k_x - cos2k_y) -    4 \sqrt{t_{2d} ^2 sin^2x sin^2y  + [\frac{t_{1s} (cosk_x + cosk_y) \pm t_{1d} (cosk_x - cosk_y)}{2}]^2},
\label{dis}
\end{eqnarray}
\end{widetext}
where $\epsilon_k=4 t_{2s} cosk_x cosk_y + 2 t_{3s} (cos2k_x + cos2k_y) - \mu$.

\begin{figure}[tbp]
\begin{center}
\includegraphics[width=1.0\linewidth]{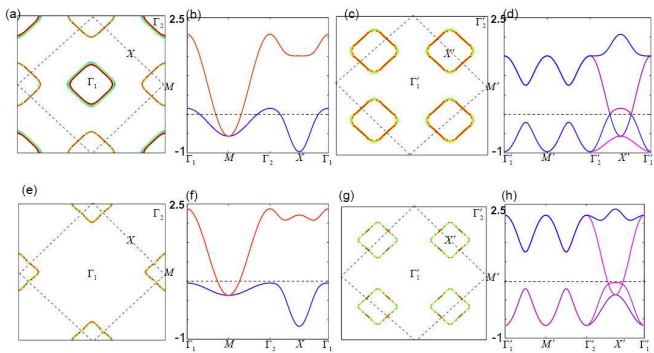}
\end{center}
\caption{  The Fermi surfaces of each component when only parameters $t_{1s}, t_{2d}$ and $t_{2s}$ are considered. The layout exactly follows Fig.\ref{fig2}. The parameters are $t_{1}=0.24$, $t_{2}=0.52$ and $\mu=-0.273$.
 The only different parameter between (a) and (e) is $t_{2}^{\prime }$ with $t_{2}^{\prime }=-0.1$ in (a) and $t_{2}^{\prime }=-0.2$ in (e). }\label{fig4}
\end{figure}

 We find that $E_{e\pm}$ can capture the electron pockets at M points and  $E_{h\pm}$ can capture the hole pockets at $\Gamma$ points. Based on the previous physical picture, $t_{1s}$, $t_{2s}$ and  $t_{2d}$ should be the largest parameters  because they are generated through the p-orbitals. In Fig.\ref{fig4},
we show that by just keeping these three parameters, the model is already good enough to capture the main characters of the bands contributing to Fermi surfaces  in the five-orbital model. 
 After performing the same gauge mapping, this Hamiltonian,   as expected,   provides  pockets located at $X'$ as shown in Fig.\ref{fig4}.  

{\it General properties of the model:} The above model is good  enough to quantitatively describe the experimental results measured by  ARPES\cite{ding-hu,hding,hding2,lxyang,ludh,chenf}.   Although the hopping parameters are dominated by $t_{1s}, t_{2d}$ and  $t_{2s}$, other parameters can not be ignored.  For example, at the same M points, there is energy splitting between two components, which indicates the existence of a sizable $t_{1d}$.  To match the detailed dispersion of the bands, the TNN hoppings have to be included. The existence of the TNN hoppings may also provide a microscopic justification for the presence of the significant TNN AFM exchange coupling $J_3$ measured by neutron scattering in iron-chalcogenides\cite{lip2,my,hu1}. 

\begin{figure}[tbp]
\begin{center}
\includegraphics[width=1.0\linewidth]{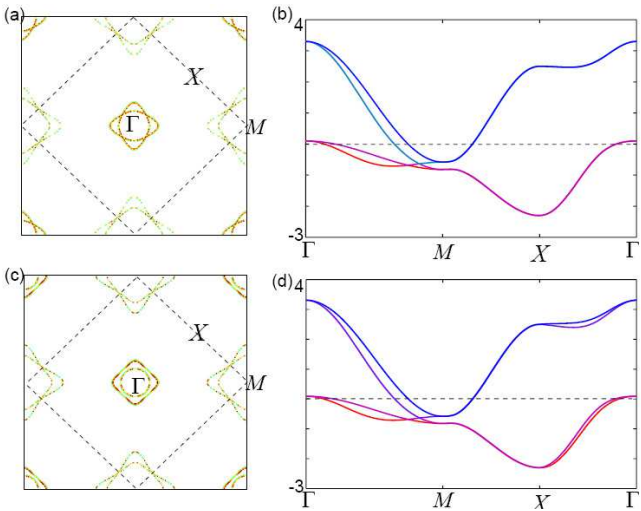}
\end{center}
\caption{  A typical Fermi surfaces (a), band dispersions (b) resulted from Eq.\ref{dis}  with parameters $t_{1s}=0.4, t_{1d}=-0.03, t_{2s}=0.3, t_{2d}=0.6, t_{3s}=0.05, t_{3d}=-0.05$ and $\mu=-0.3$. (c) and (d) are corresponding results by adding $t_c=0.02$ in Eq.\ref{tc} with the same parameter setting. }
\label{fig5}
\end{figure}

While the detailed quantitative results for different families of iron-based superconductors will be present elsewhere\cite{ding-hu},   we plot a typical case for iron-pnicitides with parameters $t_{1s}=0.4, t_{1d}=-0.03, t_{2s}=0.3, t_{2d}=0.6, t_{3s}=0.05, t_{3d}=-0.05$ and $\mu=-0.3$ in fig.\ref{fig5}(a-d). In Fig.\ref{fig5}(a,b), the coupling $t_c=0$.
In Fig.\ref{fig5}(c,d), $t_c=0.02$. It is clear that the degeneracy at the hole pockets along $\Gamma-X$ direction is lifted by $t_c$. The Fermi surfaces in Fig.\ref{fig5} are very close to those in the five orbital model\cite{kuroki}.  This result is consistent with our assumption that $t_c$ effectively is small. 

There are several interesting properties in the model.   First, the model unifies  the  iron-pnictides and  iron-chalcogenides.    When other parameters are fixed, reducing  $t_{2s}$ or increasing $t_{1s}$ can flatten the dispersion along $\Gamma-M$ direction of $E_{h\pm}$  and cause the hole pocket completely vanishes.  Therefore, the model can describe both iron-pnictides and  electron-overdoped iron-chalcogenides by varying $t_{2s}$ or $t_{1s}$. 

Second, carefully examining the hopping parameters,  we also find that  the  NNN hopping for each $S_4$ iso-spin  essentially has a $d$-wave symmetry, namely $|t_{2d}|>t_{2s}$.  Since the hole pockets can be suppressed  by reducing $t_{2s}$ value, this $d$-wave hopping symmetry is expected to be stronger in  iron chalcogenides than in iron-pnictides.  

Third, it is  interesting to point out  that  we can make an exact analogy between the $S_4$ transformation on its two iso-spin components  and  the time reversal symmetry  transformation on a real $1/2$-spin  because $S_4^2=-1$. This analogy suggests in this $S_4$ symmetric model, the degeneracy at high symmetric points in Brillouin zone is  the type of the Kramers  degeneracy. 

Finally, in this model,   if the orbital degree of freedom is included, the true unit cell for each iso-spin component includes four irons.    The gauge mapping in the previous section exactly takes a unit cell with four iron sites.   Such a match is the essential reason why the low energy physics becomes transparent after the gauge mapping.

{\it The two-orbital model with interactions:} By projecting all interactions into these two effective orbital model,  a general  effective model that describes iron-based superconductors obeying the $S_4$ symmetry can be written as  
\begin{eqnarray}
& & H_{eff}=H_{0,eff}+U\sum_{i,\alpha=1,2} \hat n_{i,\alpha\uparrow}\hat n_{i,\alpha\downarrow}\nonumber\\ 
& & +U'\sum_{i}\hat n_{i,1}\hat n_{i,2}+J_H'\sum_i \hat S_{i,1}\cdot \hat S_{i,2}
\end{eqnarray}
where $\alpha=1,2$ labels the $S_4$ iso-spin,  $U$ describes the effective Hubbard repulsion interaction within each component, $U'$ describes the one between them and $J'_H$ describes the effective Hunds coupling. Since the two components couple weakly, we may expect $U$ dominates over $U'$ and $J_H'$.  Then, in  the first order  approximation, the model could become a  single ban$d$-Hubbard model near half filling.  A similar t-J model can also be discussed within the same context  as cuprates\cite{anderson,leenw}. It is clear that the model naturally provides  an explanation for  the stable NNN AF exchange couplings $J_2$ observed by neutron scattering\cite{zhao1,lip2,my}  and  its dominating role in both magnetism and superconductivity\cite{hu1}.

{\it Reduction of the symmetry from $D_{2d}$ to   $S_4$  }
The true lattice symmetry in a $Fe-As(Se)$ trilayer is the $D_{2d}$ point group, where $S_4$ is a subgroup of the $D_{2d}$. In the $D_{2d}$ group, besides the $S_4$ invariance,  the reflection operator $\sigma_{v}$ with respect to the $x'-z$ plane is also  invariant.  The reflection imposes an additional requirement
\begin{eqnarray}\label{sigmav}
\left( \begin{array}{c}
\hat c_{k\sigma}\\
\hat d_{k\sigma}
\end{array}\right) \rightarrow \left( \begin{array}{c}
\hat c_{k^{\prime\prime}+Q\sigma}\\
-\hat d_{k^{\prime\prime}+Q\sigma}
\end{array}\right),
\end{eqnarray}   
where $k^{\prime\prime}=(k_y,k_x)$.
It is  easy to see that if we force the $D_{2d}$ symmetry, the reflection $\sigma_v$ invariance  requires $t_{1s}=0$. However, without  such a reflection invariance, this term is allowed, which is the case when only the $S_4$ symmetry remains.  

The existence of $t_{1s}$ suggests that  $\sigma_v$ symmetry must be broken in an effective model.  However, since $\sigma_v$ symmetry appears to be present, it is natural to ask  what  mechanism can break $\sigma_v$. While a detailed study of this symmetry breaking is in preparation\cite{note3}, we give a brief analysis.  Among the five d-orbitals, $d_{xy}, d_{x^2-y^2}$ and  $d_{z^2}$ belong to one dimensional representations of the $D_{2d}$ group. In fact, for these three orbitals, the $D_{2d}$ group is equivalent to $C_{4v}$ group. In other words, the As(Se) separation along c-axis has no effect on the symmetry of the kinematics of the three orbitals if the couplings to the other two orbitals, $d_{xz}$ and $d_{yz}$, are not included.  Therefore,  for these three orbitals, the unit cell is not doubled by As(Se) atoms and  the band structure is intrinsically one iron per unit cell even if the hoppings generated through p-orbitals of $As(Se)$ are important. However, for $d_{xz}$ and $d_{yz}$ orbitals, if the hoppings through p-orbitals of $As(Se)$ are dominant, the  unit cell is doubled by $As(Se)$ atoms and the band structure is intrinsically folded. From Eq.\ref{sigmav}, after the $S_4$ symmetry is maintained, the $\sigma_v$ symmetry operations  simply map the reduced Brillouin zone to the folded part.  If  the couplings between the above two groups of orbitals are turned on,  the effective two orbitals that describe the low energy physics near Fermi surfaces are not pure $d_{xz}$, $d_{yz}$ orbitals any more. In particular, they are heavily dressed by $d_{xy}$ orbitals as shown in ARPES\cite{arpes1,arpes2,arpes3,arpes4}.  Therefore, the effective two orbitals can  only keep the $S^4$ symmetry and the $\sigma_v$ symmetry has to be broken.  

Another possibility of the generation of the $t_{1s}$ hopping may stem from the following virtual hopping processes: one electron first hops from the $p_x$   to the $d_{xz}$,  then, an electron in the $p_y$  at the same As(Se) site can hop to the $p_x$, finally, an electron in the $d_{yz}$ orbital hops to the $p_y$.  In such a process, the reflection symmetry is broken due to the existence of the hopping between  the $p_x$ and $p_y$ orbitals at the same As(Se) site when the two orbitals host total 3 electrons, which is possible if onsite Hubbard interaction $U$ in $p$ orbitals is large so that the degeneracy of $p_x$ and $p_y$ is broken, a result of the  standard Jahn-Teller effect.

\begin{figure}[tbp]
\begin{center}
\includegraphics[width=1.0\linewidth]{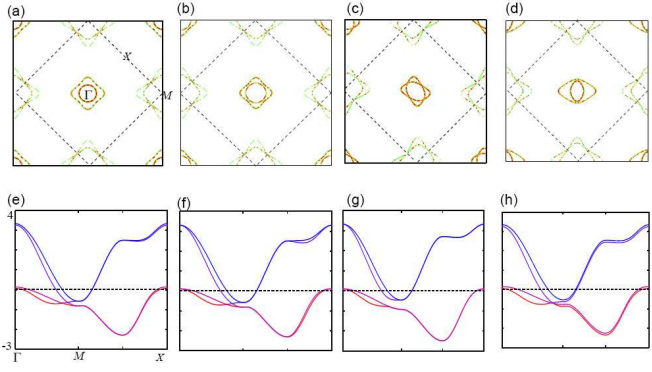}
\end{center}
\caption{  Fermi surfaces and band dispersions in the presence of the $S_4$ symmetry breaking:
(a, e)  $t_{b1}=0.005$ in Eq. \ref{b1}; (b,f) $t_{bt}=0.05$ in Eq. \ref{bt}; (c, g)  $t_{bo}=0.05$ in Eq. \ref{bo}; (d,h) $t_{bso}=0.05$ in Eq. \ref{bso}. Other parameters are the same as in Fig.\ref{fig5}. }\label{fig6}
\end{figure}
{\it The coupling between two $S_4$ iso-spins and $S_4$ symmetry breaking:} The couplings between the two iso-spins can either  keep  the $S_4$ symmetry or break it.   
Without breaking the translational symmetry,  the coupling between two orbitals can be written as 
\begin{eqnarray}
\hat H_c=\sum_{k,\alpha} f_{\alpha}(k) \hat G_{\alpha} (k)+\sum_{k,\bar\alpha} f_{\bar\alpha}(k) \hat G_{\bar\alpha} (k)
\end{eqnarray}
where   $G_{\alpha}(k)$ and $G_{\bar \alpha}(k)$ are operators constructed according to the $S_4$ one dimensional representations as follows,
\begin{eqnarray}
& & G_{1}(k)= \sum_{\sigma} c^+_{k\sigma}\hat d_{k\sigma}+c^+_{k+Q\sigma}\hat d_{k+Q\sigma} +h.c. \\
& & G_{2}(k)= \sum_{\sigma} c^+_{k\sigma}\hat d_{k\sigma}-c^+_{k+Q\sigma}\hat d_{k+Q\sigma} +h.c. \\
& & G_{3}(k)= \sum_{\sigma} c^+_{k\sigma}\hat d_{k+Q\sigma}+c^+_{k+Q\sigma}\hat d_{k\sigma} +h.c. \\
& & G_{4}(k)= \sum_{\sigma} c^+_{k\sigma}\hat d_{k+Q\sigma}-c^+_{k+Q\sigma}\hat d_{k\sigma} +h.c.\\
& & G_{\bar 1}(k)= \sum_{\sigma} i(c^+_{k\sigma}\hat d_{k\sigma}+c^+_{k+Q\sigma}\hat d_{k+Q\sigma} -h.c.) \\
& & G_{\bar 2}(k)= \sum_{\sigma} i(c^+_{k\sigma}\hat d_{k\sigma}-c^+_{k+Q\sigma}\hat d_{k+Q\sigma} -h.c.) \\
& & G_{\bar 3}(k)= \sum_{\sigma} i(c^+_{k\sigma}\hat d_{k+Q\sigma}+c^+_{k+Q\sigma}\hat d_{k\sigma} -h.c.) \\
& & G_{\bar 4}(k)= \sum_{\sigma} i(c^+_{k\sigma}\hat d_{k+Q\sigma}-c^+_{k+Q\sigma}\hat d_{k\sigma} -h.c.)
\end{eqnarray}
We discuss a few examples that can cause the $S_4$ symmetry breaking,
\begin{eqnarray}
& & H_{b1}=\sum_k 2 t_{b1} (cosk_x+cosk_y)(\hat c^+_{k\sigma}\hat d_{k+Q\sigma}+h.c.) 
\label{b1} \\
& & H_{bt}=\sum_k4i t_{bt} sink_xsink_y (\hat c^+_{k\sigma}\hat d_{k+Q\sigma}-h.c.)
\label{bt}  \\
& & H_{bo}=\sum_k t_{bo}(\hat c^+_{k\sigma}\hat c_{k+Q\sigma}-d^+_{k\sigma}\hat d_{k+Q\sigma})
\label{bo}\\  
& & H_{bso}=\sum_k t_{bso}(\hat c^+_{k\sigma}\hat c_{k\sigma}-d^+_{k\sigma}\hat d_{k\sigma}).
\label{bso}
\end{eqnarray}$t_{b1}$ term breaks the $S_4$ symmetry to lift the degeneracy at $\Gamma$ point, $t_{bt}$ breaks  the time reversal symmetry, $t_{bo}$ indicates a ferro-orbital ordering and $t_{bso}$ indicates a staggered orbital ordering. These terms can be generated either spontaneously or externally and their effects can be explicitly observed in the change of the band structure and degeneracy lifting as shown  in Fig.\ref{fig6}, where   the changes of  band structures and Fermi surfaces   due to the symmetry breaking terms are plotted.   It is fascinating to study how the interplay between the  $S_4$ symmetry and other broken symmetries in this system in future.

\section{The classification of the superconducting orders according to the $S_4$ symmetry}The presence of the $S_4$ symmetry brings a new symmetry classification of the superconducting phases.  The $S_4$ point group has four one-dimensional representations, including $A$, $B $ and $2E$. In the $A$ state, the $S_4$ symmetry is maintained.  In the B state, the state changes sign under the $S_4$ transformation.  In the $2E$ state, the state obtains a $\pm \pi/2$ phase under the $S_4$ transformation.
Therefore, the $2E$ state breaks the $C_2$ rotational symmetry as well as the time reversal symmetry.
  
Since the $S_4$ transformation includes two parts, a $90^\circ$ degree rotation and a reflection along c-axis, the $S_4$ symmetry classification leads to a natural correlation between the rotation in a-b plane and c-axis reflection symmetries in a SC state.  In the A-phase,  rotation and c-axis reflection can be both broken,  while in the B-phase, one and only one of them can be broken. This correlation, in principle, may be  observed  by applying external symmetry breaking. For example, even in the A-phase and the rotation symmetry is not broken, we may force the c-axis phase-flip to obtain the phase change in the a-b plane.  

As shown in this paper, the iron-based superconductors are rather unique with respect to the $S_4$ symmetry. It has two iso-spin components governed by the symmetry. This iso-spin  degree of freedom and the interaction between them  could lead to many novel phases.  The future study can explore these possibilities. 

Here we specifically discuss  the $S_4$ symmetry aspects  in the proposed $A_{1g}$ $s$-wave state, a most-likely phase  if it is driven by the   repulsive interaction or strong AF in iron-based superconductors\cite{seo2008} as we have shown earlier. First, let's clarify the terminology issues. The $A_{1g}$  $s$-wave pairing symmetry  is classified according to $D_{4h}$ point group. This classification is not right in the view of the true lattice symmetry. However,  for each iso-spin components, we can still use it.  Here we treat it as a state that the superconducting order $\Delta\propto cosk_xcosk_y$\cite{seo2008}.   Since the $A_{1g}$ phase is equivalent to the $d$-wave in cuprates in a different gauge setting, the $d$-wave picture is more transparent regarding the sign change of the superconducting phase in the real space.  As shown in Fig.\ref{fig1}, the sign of the SC order alternates  between neighboring squares in the iron lattice.

\begin{figure}[tbp]
\begin{center}
\includegraphics[width=1.0\linewidth]{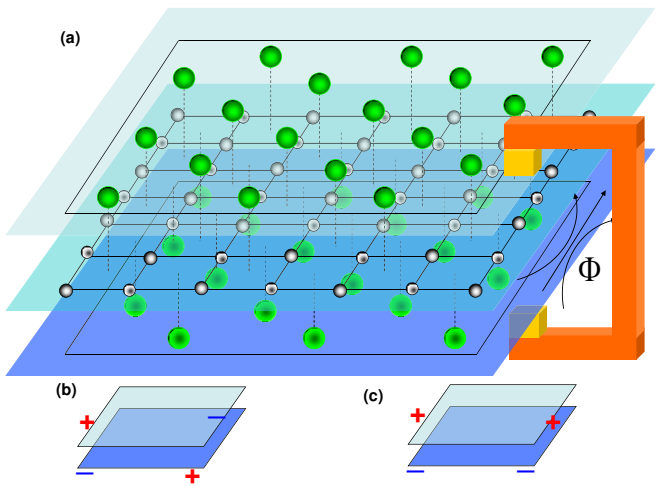}
\end{center}
\caption{ (a) An illustration of a single Fe-As(Se) layer and  the setup for a dc SQUIDS measurement to measure the sign change of the SC phase between top and down As(Se) layers; (b) The phase distribution in the  A phase  of the $A_{1g}$ $s$-wave state in the view of a $d$-wave picture ( red for one iso-spin component and blue for the other); (c) The phase distribution in the   B phase  of the $A_{1g}$ $s$-wave state.}\label{fig7}
\end{figure}
Based on the underlining electronic structure revealed here,  with respect to the $S_4$ symmetry, the $A_{1g}$ state can have  two different phases,  A phase and   B phase.  In the A phase,
\begin{eqnarray}
<\hat c_{k\uparrow} \hat c_{-k\downarrow}>=<\hat d_{k\uparrow} \hat d_{-k\downarrow}>=\Delta_0 cosk_xcosk_y
\end{eqnarray}
and in the B phase, 
\begin{eqnarray}
<\hat c_{k\uparrow} \hat c_{-k\downarrow}>=-<\hat d_{k\uparrow} \hat d_{-k\downarrow}>=\Delta_0 cosk_xcosk_y.
\end{eqnarray}
Therefore, in the view of the $d$-wave picture,    in both A and B phases, the superconducting phase  for each component alternates between neighboring squares, which is corresponding to the sign change  between the top and bottom planes  in the view of the $S_4$ symmetry. However, in the A phase,  since the $S_4$ symmetry is not violated, the relative phase between the two components are equal to $\pi$ in space, while in the B phase, the relative phase   is zero.  A picture of the phase distribution of the two iso-spin components in the A and B phases are illustrated in Fig.\ref{fig7}(b,c). 

 The sign change of the order parameter or the phase shift of $\pi$  between the top and bottom planes along c-axis can be detected by standard magnetic flux modulation of dc SQUIDS measurements\cite{woll}.  If we consider a single Fe-As(Se) trilayer structure, which has been successfully grown by MBE recently\cite{liud,qwang},  we can design a standard dc SQUIDS as shown in Fig.\ref{fig7}(a) following the similar experimental setup to determine  the $d$-wave pairing in cuprates\cite{woll}. For the B phase, there is no question that the design can repeat the previous results in cuprates.  However, if the tunneling matrix elements to two components are not symmetric, even in the A phase, this design can obtain the signal of the  $\pi$ phase shift  since the two components are weakly coupled and each of them has a $\pi$ phase shift.  For the B phase,  the phase shift   may be preserved even in bulk materials\cite{note}. However, for the A phase,  it  will be difficult  to detect the phase shift in bulk materials.  A cleverer design is needed.  Measuring the phase shift between the upper and lower As(Se)  planes will be a smoking-gun experiment to verify the model and determine iron-based superconductors and cuprates sharing identical superconducting mechanism.

\section{Discussion and Summary}

\begin{figure}[tbp]
\begin{center}
\includegraphics[width=1.0\linewidth]{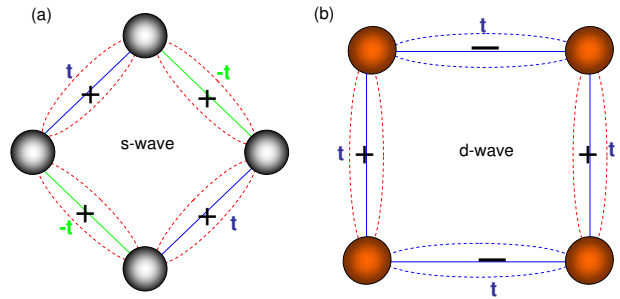}
\end{center}
\caption{ A sketch of the correlation between the hopping  and pairing symmetries for both iron-based superconductors and cuprates. }\label{fig8}
\end{figure}
We have shown that the $A_{1g}$ $s$-wave pairing in iron-based superconductors is  a $d$-wave pairing in a view of a different gauge setting. This equivalence  answers an essential question why  a $A_{1g}$ $s$-wave pairing  can be robust  regardless of the presence or absence of the hole pockets.   With repulsive interactions, a sign changed order parameter in a superconducting state is usually inevitable. This statement, however, is only true when the hopping parameters 
follow the same lattice symmetry. Gauge transformation can exchange   the phases between superconducting order parameters and hopping parameters.  In the case of cuprates, the $d$-wave order parameter can be transformed to a $s$-wave form by changing hopping parameters to  obey  $d$-wave symmetry.   As we pointed out earlier,  the NNN hopping in our model is close to a $d$-wave symmetry,  rather than a $s$-wave symmetry. This is the essential reason why the superconducting order can have a $s$-wave form and be stable in iron-based superconductors.  A simple picture of this discussion is illustrated in Fig.\ref{fig8}.  The vanishing of the hole pockets in electron-overdoped iron-chalcogenides indicates the hopping is even more $d$-wave like in these materials, a case supporting stronger $s$-wave pairing, which was indeed observed  recently\cite{lsun,qwang}.  The presence of  the dominant form $cosk_xcosk_y$ is also straightly linked to the $d$-wave pairing form $cosk_x'-cosk_y'$ because of the stable AF $J_2$ coupling, similar to cupates\cite{kotliar}.  Moreover, since the different gauge setting does not alter physical measurements, a phase sensitive measurement should reveal  a $\pi$ phase shift  in the real space along c-axis for each components in the $A_{1g}$ $s$-wave state, just like the phase shift along a and b direction in the $d$-wave pairing state of cuprates.

After obtaining the underlining electronic structure,  we can ask how the physics in the  cuprates and iron-based superconductors are related to each other.  In Table.\ref{table}, we list the close relations between two high $T_c$ superconductors. From the table, it is clear that  by determining these physical properties of iron-based superconductors listed in the table can help  to determine  the high $T_c$ superconducting mechanism. 
\begin{center}
\begin{table}[tbp]
\begin{tabular}{|c|c|c||}
\hline properties & Iron SCs & cuprates\\ \hline
  pairing symmetry &  $s$-wave & $d$-wave  \\ \hline
underlining hopping symmetry & $d$-wave  & $s$-wave  \\ \hline
dominant pairing form & $cosk_xcosk_y$ & $cosk_x-cosk_y$  \\ \hline
pairing classification symmetry &  $S_4$ & $C_4$  \\ \hline
AF coupling & NNN $J_2$  & NN $J_1$  \\ \hline
sign change in real space & c-axis  & a-b plane   \\ \hline
filling density  &  half-filling  & half-filling \\ \hline
\end{tabular}
\caption{ A list of the close connections between iron-based superconductors (iron SCs)  and  cuprates.}
\label{table}
\end{table}
\end{center}

The model completely changes the view of the origin of the  generation of sign-changed $s^\pm$ pairing symmetry in iron-pnictides, which were argued in many theories that the origin is the scattering between electron pockets at $M$ and hole pockets at $\Gamma$ due to repulsive interactions\cite{hirschfeld, mazin}.  With the new underlining electronic structure revealed,  the analysis of the sign-change should be examined after taking the gauge transformation so that the underlining hopping parameters become symmetric.  In this case, the sign change is driven by  scatterings   between all pockets, including both hole and electron pockets,  located at two $d$-wave anti-nodal $X'$ points.   Therefore, the scattering between electron and electron pockets is also important. 

While the model appears to be rotational invariant due to the $S_4$ symmetry,  the dynamics of each iso-spin component  is intrinsically nematic.  A small $S_4$ symmetry breaking can easily lead to an overall electronic nematic state. 
The electronic nematic state has been observed  by many experimental techniques\cite{rfisher} and studied by different theoretical models\cite{cfang,cxu,huxu,mazin2,kruger,wlv,cclee,fernandez}.  The underlining electronic structure in the model can provide a straightforward microscopic understanding between the interplay of all different degree of freedoms based on the $S_4$ symmetry breaking. 

It is worth to pointing out that  in our model, if  $t_{1s}$ is generated by the mixture of different orbital characters, $t_{1s}$ is generally not limited to the NN hopping. It can be a function of $k$ which satisfies $t_{1s}(k)=-t_{1s}(k+Q)$ so that it breaks $\sigma_v$ symmetry. The value of $t_c$ may be not  small. However,  both $t_{1s}$ and $t_c$  have very limited effects on  the electron pockets.    While we may use a different set of $t_{1s}$ and $t_c$ to fit the electronic structure, the key physics in the paper remains the same because  the essential physics stems from the NNN hoppings,

In summary, we have shown the underlining electronic structure, which is responsible for superconductivity at low energy  in iron-based superconductors,  is essentially  two nearly degenerated    electronic structures governed by the $S_4$ symmetry.  We demonstrate the $s$-wave pairing in iron-based superconductors is equivalent to the $d$-wave in cuprates. A similar conclusion has also been reached in the study of 2-layer Hubbard model\cite{scalapino2}.  The $S_4$ symmetry reveals possible new superconducting states and suggests the phase shift in the SC state in real space is along c-axis.  These results strongly support the microscopic superconducting mechanism for cuprates and iron-based superconductors are identical, including both iron-pnictides and iron-chalcogenides.   Our model establishes a new foundation for understanding and exploring properties of iron based superconductors, a unique, elegant and beautiful class of superconductors. 

{\it Acknowledgement:}  JP thanks H. Ding, D.L. Feng, S. A, Kivelson, P. Coleman, X Dai, Y.P. Wang,  EA Kim and F. Wang for useful discussion. JP specially thanks H.Ding, F. Wang, M. Fischer and W. Li for  the discussion of the symmetry properties of the model. The work is supported  by the Ministry of Science and Technology of China 973 program(2012CB821400) and NSFC-1190024.
%



\begin{thebibliography}{66}
\bibitem [{\citenamefont {Note}()}]{note}%
  \BibitemOpen
  \bibfield  {author} {\bibinfo {author} {\bibnamefont {Note}},\ }\href@noop {}
  {}\bibinfo {note} {The 122 structure doubles unit cell along c-axis. If it is
  in the B-phase, it requires odd number of iron-layers to measure the sign
  change. We will discuss this issue in the future.}\BibitemShut {Stop}%
\end{thebibliography}
\end{document}